\documentclass[12pt,psfig]{article}
\makeatletter
\def\@cite#1#2{{[{#1}]\if@tempswa\typeout
{IJCGA warning: optional citation argument
ignored: `#2'} \fi}}

\newcount\@tempcntc
\def\@citex[#1]#2{\if@filesw\immediate\write\@auxout{\string\citation{#2}}\fi
  \@tempcnta\z@\@tempcntb\m@ne\def\@citea{}\@cite{\@for\@citeb:=#2\do
    {\@ifundefined
       {b@\@citeb}{\@citeo\@tempcntb\m@ne\@citea\def\@citea{,}{\bf ?}\@warning
       {Citation `\@citeb' on page \thepage \space undefined}}%
    {\setbox\z@\hbox{\global\@tempcntc0\csname b@\@citeb\endcsname\relax}%
     \ifnum\@tempcntc=\z@ \@citeo\@tempcntb\m@ne
       \@citea\def\@citea{,}\hbox{\csname b@\@citeb\endcsname}%
     \else
      \advance\@tempcntb\@ne
      \ifnum\@tempcntb=\@tempcntc
      \else\advance\@tempcntb\m@ne\@citeo
      \@tempcnta\@tempcntc\@tempcntb\@tempcntc\fi\fi}}\@citeo}{#1}}
\def\@citeo{\ifnum\@tempcnta>\@tempcntb\else\@citea\def\@citea{,}%
  \ifnum\@tempcnta=\@tempcntb\the\@tempcnta\else
   {\advance\@tempcnta\@ne\ifnum\@tempcnta=\@tempcntb \else
\def\@citea{--}\fi
    \advance\@tempcnta\m@ne\the\@tempcnta\@citea\the\@tempcntb}\fi\fi}
\makeatother

\newcommand{\gsim}{\lower.7ex\hbox{$\;\stackrel{\textstyle>}{\sim}\;$}}
\newcommand{\lsim}{\lower.7ex\hbox{$\;\stackrel{\textstyle<}{\sim}\;$}}
\newcommand{\be}{\begin{equation}}
\newcommand{\ee}{\end{equation}}
\newcommand{\bea}{\begin{eqnarray}}
\newcommand{\eea}{\end{eqnarray}}

\def\baselinestretch{1}
\begin{document}
\catcode`@=11
\newtoks\@stequation
\def\subequations{\refstepcounter{equation}%
\edef\@savedequation{\the\c@equation}%
  \@stequation=\expandafter{\theequation}
  \edef\@savedtheequation{\the\@stequation}
  \edef\oldtheequation{\theequation}%
  \setcounter{equation}{0}%
  \def\theequation{\oldtheequation\alph{equation}}}
\def\endsubequations{\setcounter{equation}{\@savedequation}%
  \@stequation=\expandafter{\@savedtheequation}%
  \edef\theequation{\the\@stequation}\global\@ignoretrue

\noindent}
\catcode`@=12
\begin{titlepage}

\title{{\bf
Spheres, Deficit Angles and the Cosmological Constant}} \vskip2in
\author{
{\bf Ignacio Navarro$$\footnote{\baselineskip=16pt E-mail: {\tt
ignacio.navarro@durham.ac.uk}}}
\hspace{3cm}\\
 $$~{\small IPPP, University of Durham, DH1 3LE Durham, UK}.
}
\date{}
\maketitle
\def\baselinestretch{1.15}
\begin{abstract}
\noindent We consider compactifications of six dimensional gravity
in four dimensional Minkowski or de Sitter space times a two
dimensional sphere, $S^2$. As has been recently pointed out, it is
possible to introduce 3-branes in these backgrounds with arbitrary
tension without affecting the effective four dimensional
cosmological constant, since its only effect is to induce a
deficit angle in the sphere. We show that if a monopole like
configuration of a 6D $U(1)$ gauge field is used to produce the
spontaneous compactification of the two extra dimensions in a
sphere a fine tuning between brane and bulk parameters is
reintroduced once the quantization condition for the gauge field
is taken into account, so the 4D cosmological constant $depends$
on the brane tension. This problem is absent if instead of the
monopole we consider a four form field strength in the bulk to
obtain the required energy-momentum tensor. Also, making use of
the four form field, we generalize the solution to an arbitrary
number of dimensions ($\ge 6$), keeping always four noncompact
dimensions and compactifying the rest in a $n$-dimensional sphere.
We show that a ($n$+1)-brane with arbitrary tension can be
introduced in this background without affecting the effective 4D
cosmological constant.

\end{abstract}

\thispagestyle{empty} \vspace{5cm}  \leftline{}

\vskip-24cm \rightline{} \rightline{IPPP/03/20}
\rightline{DCPT/03/40} \vskip3in

\end{titlepage}
\setcounter{footnote}{0} \setcounter{page}{1}
\newpage
\baselineskip=20pt

\section{Introduction}

\noindent The old idea that the universe has more than four
dimensions has been the subject of intense research activity in
recent years, due mainly to the observation that some fields can
be confined in a submanifold of the total space. This has opened
new avenues in (brane world) model building and some of the major
problems in theoretical physics have been addressed in this
fashion with very interesting results. For instance, regarding the
hierarchy problem, in the models of
\cite{Arkani-Hamed:1998rs,Randall:1999ee} it has been shown how
the scale of quantum gravity can become a free parameter. Theories
in which Standard Model fields are confined in a brane embedded in
a higher dimensional bulk space have also been used to attack the
cosmological constant problem. In
\cite{Arkani-Hamed:2000eg,Kachru:2000hf}, solutions were found in
a 5D model where a 3-brane is kept flat for any value of it´s
tension making use of a scalar field with particular couplings in
the bulk and to the brane tension. However, this kind of models
always involve a naked spacetime singularity at finite proper
distance from the brane \cite{Csaki:2000wz}, and the resolution of
this singularity in a more fundamental theory of gravity can
reintroduce a fine tuning between bulk parameters and the brane
tension \cite{Forste:2000ft}. Recently a similar property was
found in 6D models \cite{Carroll:2003db,Navarro:2003vw} that are
free of singularities\footnote{In \cite{Chen:2000at}, it was
noticed that the only effect of a three brane in a 6D bulk that is
rotationally symmetric around the brane axis would be to induce a
deficit angle in the transverse space, but the compactifications
considered always involved a spacetime singularity or did not
maintain the nice property that the effective cosmological
constant in 4D was independent of the brane tension. In contrast,
the models of \cite{Carroll:2003db,Navarro:2003vw} do not involve
any spacetime singularities except for a conical one that can be
easily smoothed out if we allow for a finite brane thickness.}.
These models have two extra compact dimensions and it has been
shown that it is possible to introduce three branes in this
background with arbitrary tensions and the effective 4D
cosmological constant is completely independent these brane
tensions, only depending on bulk parameters. The only effect of
the branes is to introduce a deficit angle in the compactification
manifold (taken to be a sphere or a disk). This opens the
possibility of building models in which the required fine tunings
are the consequence of some unbroken symmetry in the bulk ($e.g.$
supersymmetry) while this symmetry can be badly broken in the
brane without affecting the effective cosmological constant.

The cosmology and stability of the scenarios sketched above has
been studied in \cite{Cline:2003ak}, where it is argued that the
model is stable since all scalar fluctuations of the metric are
massive and only the graviphoton and the 4D graviton are massless.
Also, non standard cosmological properties are emphasized. In this
paper we will continue the exploration of these models. We will
consider different mechanisms for producing the required
spontaneous compactification of the extra dimensions, and we will
generalize the solution to an arbitrary number of dimensions. In
the next section we show that if a monopole like configuration of
a $U(1)$ gauge field is used for producing the spontaneous
compactification of two extra dimensions in a sphere as in
\cite{Carroll:2003db,Randjbar-Daemi:1982hi}, a fine tuning between
bulk and brane parameters is reintroduced once we consider the
quantization condition for the gauge field. This happens because
the quantization condition for the monopole configuration depends
on the deficit angle of the sphere, so to obtain a small enough
effective cosmological constant in 4D we must fine tune the 6D
cosmological constant against the $U(1)$ charge $and$ the brane
tension (via the deficit angle). We will show that this problem is
not present if instead of considering a two form with a vacuum
expectation value in order to generate the required
energy-momentum tensor we consider a four form with a Freund-Rubin
configuration \cite{Freund:1980xh} in the bulk. In this case the
fine tuning required in order to obtain flat 4D space does not
involve brane parameters. In section 3 we will generalize the
solution to an arbitrary number of dimensions. Making use of a
four form field we will obtain solutions in which the space is the
direct product of 4 dimensional Minkowski or de Sitter space
($dS_4$) times a $n$ dimensional sphere, $S^n$. In order to obtain
Minkowski space at low energies we have to fine tune the $d=4+n$
dimensional cosmological constant with the vacuum expectation
value (v.e.v.) of the four form field.

Again, we will be able to introduce a $(d-3)$-brane in this
background with arbitrary tension and we will see that the
effective 4D cosmological constant is independent of this tension,
since the effect of the brane is to induce a deficit angle in the
two dimensional transverse space. The tuning required in order to
obtain flat 4D space is independent of brane parameters also in
this general case, since it only involves the bulk cosmological
constant and the four form v.e.v.. The brane will have four
noncompact flat dimensions and $n-2$ dimensions compactified in a
sphere $S^{n-2}$, wrapped around the original $n$-dimensional
sphere. Section 4 is devoted to the conclusions.

\section{Compactification in $S^2$ Using Two and Four Form Fields.}
\subsection{The 6D background solution.}

We will start by briefly reviewing the basic features of the 6D
model of \cite{Navarro:2003vw}, without assuming a particular
origin for the energy-momentum tensor. The metric ansatz assumes a
factorizable geometry,

\be ds^2 = \gamma(x)_{\mu \nu}dx^{\mu}dx^{\nu}+\kappa(z)_{ij} \;
dz^idz^j \label{ansatz1}.\ee

\noindent Lower case latin indices run over the two extra
dimensions, greek indices over the four conventional ones and
upper case letters over all the coordinates. The required form of
the energy momentum tensor is

\be
T_{MN}=-\left(\begin{array}{cc}\gamma_{\mu \nu}\Lambda_{\gamma} & \\
& \kappa_{ij}\Lambda_{\kappa}\end{array} \right).\label{emt}\ee

\noindent The solution is simply the direct product of 4D
Minkowski or $dS_4$ space times a 2 sphere,

\be ds^2 = dt^2 - e^{Ct}d\vec{{\bf x}}^2 - R^2_0 (d\theta^2 +
\beta^2 sin^2\theta \;d\varphi^2) \label{sol} \ee

\noindent where $\theta$ ranges from 0 to $\pi$ and $\varphi$
takes values in $[0,2\pi)$. The constants $C$ and $R_0$ are given
by

\be C^2 = -{2\over 3} \Lambda_{\kappa} \;\; , \;\;
R^{-2}_0={1\over 2}\Lambda_{\kappa}-\Lambda_{\gamma}.\label{ctes}
\ee

\noindent In here and in the following we will take units in which
the higher dimensional Planck mass is one, so powers of this mass
should be understood where needed. If $\beta \neq 1$ there is a
conical singularity at those points where $sin(\theta)$ vanishes.
This can also be cast as a deficit angle redefining the coordinate
$\varphi$ so it ranges from 0 to $2\pi \beta$ and then the metric
for this new coordinate system would take a conventional form
(like (\ref{sol}) with $\beta=1$). The Einstein equations will be
satisfied in all the spacetime if we consider the presence of
three branes at this locations with a tension given by
\cite{Carroll:2003db,Navarro:2003vw}

\be 1-\beta={T_0\over 2\pi}.\label{defang} \ee

We see that if the compactification manifold is a sphere we need
to consider configurations with two branes of equal tensions at
antipodal points of the sphere. This could raise concerns since it
can be viewed as a fine tuning involving the brane tensions.
However, it is also possible to consider compactifications in the
disk. The disk is the orbifold $S^2/Z_2$, obtained from the sphere
under the identification of points in the northern and southern
hemispheres that are symmetric under reflections through the
equator. In this case no second three brane is needed, and it was
shown in \cite{Navarro:2003vw} that the matching conditions in the
orbifold fixed line (the equator) are trivially satisfied, so we
do not need to consider the presence of any four brane at these
points that could reintroduce a fine tuning between brane and
off-brane parameters \cite{Chen:2000at}. The interesting feature
of this model is that the fine tuning required to obtain flat 4D
space ($\Lambda_{\kappa}=0$, as can be seen from eq.(\ref{ctes}))
does not involve brane parameters, and we can find solutions with
flat 4D space for any value of the brane tension once we have made
this tuning. The only effect of the brane is to induce a deficit
angle in the sphere, without affecting the effective 4D
cosmological constant. Supersymmetry is probably the best
candidate for ensuring that the bulk theory provides an
energy-momentum tensor with $\Lambda_{\kappa}\simeq 0$ after
including radiative corrections if it is (approximately) preserved
in the bulk, like in the supergravity models of
\cite{Salam:1984cj,Aghababaie:2002be}. In these models flat 4D
space is a direct consequence of the SUSY 6D Lagrangian, with no
fine tuning involved. They use a monopole configuration to
compactify two extra dimensions in a sphere, and N=1 supersymmetry
is preserved by the compactification. However, once we introduce a
brane with nonzero tension, the flat 4D solution is no longer
preserved for reasons similar to those explained in the next
subsection, since the monopole configuration is affected by the
deficit angle induced by the brane. If another 6D supersymmetric
model is found that produces naturally flat 4D space before
supersymmetry breaking, and the solution is not affected by a
deficit angle in the transverse space, supersymmetry could be
broken in the brane without affecting the effective cosmological
constant\footnote{In any complete model one should of course
compute the transmission of supersymmetry breaking from the brane
to the bulk and make sure that this symmetry breaking effects do
not produce a cosmological constant in 4D that is too large.}.

\subsection{Monopole vs. Four Form}

In the previous section, as well as in \cite{Navarro:2003vw}, we
did not assume a particular mechanism for generating the required
v.e.v. of the energy-momentum tensor, eq.(\ref{emt}). Since we are
decomposing the total 6D space in 4+2 dimensions, following
\cite{Freund:1980xh}, a natural candidate for generating this
inhomogeneity would be the v.e.v. of a 2-form or a 4-form field
strength. In \cite{Carroll:2003db,Randjbar-Daemi:1982hi} the two
form option was chosen, and the resulting configuration is that of
a magnetic monopole field. Adding a bulk cosmological constant and
tuning it´s  value against the two form v.e.v. one can find flat
4D solutions. Here we will show that once the quantization
condition is imposed on the monopole field, since the 2-form
v.e.v. depends on the deficit angle, the required tuning involves
the brane tension, so the effective 4D cosmological constant
$does$ depend on brane parameters, spoiling the nice features of
the model. This problem can be overcome using a 4-form field with
a v.e.v. that will not depend on the deficit angle of the
transverse space.

\subsubsection{Monopole.}

The Lagrangian in this case is that of Einstein-Maxwell theory
with a cosmological constant

\be S = \int d^6 x \sqrt{-g}\left(-{1\over 2}R - {1\over
4}F_{MN}F^{MN} -
 \Lambda \right)\label{lagrangianEMT}.\ee

\noindent The metric (\ref{sol}) will be a solution to the coupled
Einstein-Maxwell equations in this theory if the assumed v.e.v. of
the $U(1)$ gauge field is that of a magnetic monopole
\cite{Randjbar-Daemi:1982hi},

\be A_{M}dx^M = B (cos(\theta) \pm 1 ) d \varphi \label{gauge},\ee

\noindent with $F_{MN}=\partial_M A_N - \partial_N A_M$. The plus
and minus signs are needed to cover the upper and lower
hemispheres of the sphere so the vector field is well defined
everywhere. The necessity of a quantization of the constant $B$
can be seen by requiring that a single valued gauge transformation
relates the two representations of the gauge field where they
overlap, in the equator:

\be A_{\varphi}^+ = A_{\varphi}^- + \partial_\varphi
\alpha(\varphi)\ee

\noindent so $\alpha(\varphi) = 2 B\varphi$. Now imposing that the
gauge transformation $e^{ig\alpha(\varphi)}$ is single valued we
get \cite{Randjbar-Daemi:1982hi}

\be B = {n\over 2 g} \label{cuant}\ee

\noindent with $n$ an integer. The energy-momentum tensor we get
from the action (\ref{lagrangianEMT}) with the ansatz
(\ref{gauge}) is

\be T_{MN} =  \left(\begin{array}{cc}\gamma_{\mu \nu}\left( {n^2\over 8g^2\beta^2 R_0^4} + \Lambda\right) & \\
& \kappa_{ij}\left( -{n^2\over 8g^2\beta^2 R_0^4} +
\Lambda\right)\end{array} \right).\label{EMT2}\ee

\noindent In order to obtain flat 4D space we must fine tune the
coefficient of $\kappa_{ij}$ above to zero, so the required fine
tuning involves the deficit angle\footnote{We could also have used
a coordinate system in which the parameter $\beta$ does not appear
in the metric, but the maximum allowed value for $\varphi$ is
$2\pi \beta$. In this case the factor of $\beta$ in
eq.(\ref{EMT2}) would not come from the metric, but from the
quantization condition eq.(\ref{cuant}), that now would have a
factor of $1/\beta$ in the right hand side.}. Once this tuning is
made for a value of the brane tension, any other value of this
tension would generate an effective cosmological constant. The
reason for this is clear: the monopole configuration knows about
the deficit angle of the sphere since it obeys a topological
quantization condition, and it reacts under changes in this
parameter. So if we want to maintain the nice property that the
effective 4D vacuum energy is independent of the brane tension we
need another mechanism for generating an inhomogeneous energy
momentum tensor.

\subsubsection{Four Form Field Strength.}

The difficulties we found in the previous construction can be
overcome if we consider a four form field with a Freund-Rubin
configuration \cite{Freund:1980xh} instead of the monopole. Four
form fields appear naturally in supergravity theories and have
been used in work related to the cosmological constant problem
\cite{Turok:1998he,Bousso:2000xa}. The action we take in this case
is

\be S = \int d^6 x \sqrt{-g}\left(-{1\over 2}R - {1\over
48}F_{MNPQ}F^{MNPQ} -  \Lambda \right)\label{lagrangian4F6D}.\ee

\noindent Again, the metric (\ref{sol}) is a solution of the field
equations if the four form has a v.e.v. given by

\be F^{\mu \nu \lambda \sigma} = {E \over
\sqrt{-\gamma}}\;\;\epsilon^{\mu \nu \lambda \sigma},
\label{4fa}\ee

\noindent where $\epsilon^{\mu \nu \lambda \sigma}$ is the totally
antisymmetric tensor, $E$ is an arbitrary constant\footnote{In
\cite{Bousso:2000xa} it was argued that this flux has a
quantization condition once we embed this construction in M
theory. Whether this flux is quantized or not is not important for
our purposes, since in any case the quantization condition would
be independent of the deficit angle.} and indices run over the
four conventional dimensions only. The rest of the elements of
$F^{MNPQ}$ are zero. This ansatz solves the four form field
equations and yields the following energy-momentum tensor

\be T_{MN} =  \left(\begin{array}{cc}\gamma_{\mu \nu}\left( {E^2\over 2} + \Lambda\right) & \\
& \kappa_{ij}\left( -{E^2\over 2} + \Lambda\right)\end{array}
\right).\ee

\noindent By comparing this with the formulas of subsection 2.1 we
see that the tuning required to obtain flat 4D space is now
$E\simeq \sqrt{2\Lambda}$, that is independent of the deficit
angle. Some mechanism is required to ensure this cancellation and
its stability under radiative corrections to fully address the
cosmological constant problem (as we have already commented a good
candidate could be supersymmetric models perhaps along the lines
of \cite{Salam:1984cj,Aghababaie:2002be}). The important point
now, and the reason why these models can be regarded as progress
in finding an eventual solution to this problem is that once this
tuning is made, the cosmological constant is zero for any value of
the brane tension.

\section{Generalizing to d=4+n dimensions.}

The four form has also another advantage with respect to the
monopole configuration, and it is that the solution can be easily
generalized to an arbitrary number of dimensions, keeping always
four noncompact dimensions, and compactifying the rest in a
$n$-dimensional sphere. The reason for this is that the
Freund-Rubin configuration we are assuming for the four form
distinguishes four dimensions from the rest and provides an
inhomogeneous energy-momentum tensor that differentiates these
four dimensions. In this way we can find solutions of the
$d$-dimensional Einstein equations with four flat dimensions,
while the rest are compactified in a manifold of constant
curvature, like a sphere. In this background we will be able to
introduce a $(n+1)$-brane with arbitrary tension, without
affecting the 4D cosmological constant, since its only effect will
be to produce a decifit angle in $S^n$. For the deficit angle
mechanism to work we need a codimension two brane, so some of the
dimensions of the brane will be compact, forming a
$(n-2)$-dimensional sphere that wraps around the original
$n$-dimensional sphere.

\subsection{The Background Solution.}

The starting point here will be the action

\be S = \int d^d x \sqrt{-g}\left(-{1\over 2}R - {1\over
48}F_{MNPQ}F^{MNPQ} - \Lambda \right)\label{lagrangian4FnD},\ee

\noindent where $d=4+n$. We will look for solutions in which the
total space is the direct product of two Einstein manifolds: 4D
Minkowski or $dS_4$ times $S^n$. The metric ansatz is again

\be ds^2 = \gamma(x)_{\mu \nu}dx^{\mu}dx^{\nu}+\kappa(z)_{ij} \;
dz^idz^j \label{ansatz2},\ee

\noindent where now the lower case latin indices run over the $n$
extra dimensions. The ansatz for the four form field is as in
eq.(\ref{4fa}), and it can be easily seen that is solves the four
form equation of motion \cite{Freund:1980xh}. Now Einstein
equations are

\be -\left(\begin{array}{cc}\gamma_{\mu \nu}\left( {1\over 4}R(\gamma) + {1 \over 2}R(\kappa)\right) & \\
& \kappa_{ij}\left( {1\over 2}R(\gamma) + {n-2\over 2n}R(\kappa)
\right)\end{array} \right)=
\left(\begin{array}{cc}\gamma_{\mu \nu}\left( {E^2\over 2} + \Lambda\right) & \\
& \kappa_{ij}\left( -{E^2\over 2} + \Lambda\right)\end{array}
\right).\label{Eeqs}\ee

\noindent $R(\gamma)$ and $R(\kappa)$ are the curvatures of the
respective submanifolds. It is easy to solve for this curvatures,
we get

\be R(\gamma) = -{8\over n+2}\Lambda + 4 {n-1\over n+2}E^2
\;\;\;,\;\;\; R(\kappa) = -{2n\over n+2}\left(\Lambda + {3\over
2}E^2 \right).\ee

\noindent We can write the metric in the form

\be ds^2 = dt^2 - e^{Ct}d\vec{{\bf x}}^2 - R^2_0 d\Omega_n^2
\label{sol-n-dim} \ee

\noindent where $d\Omega_n^2$ is defined as

\be d\Omega_n^2 = d\theta^2_{n-1} + sin^2 \theta_{n-1}
d\Omega_{n-1}^2 \ee

\noindent with $d\Omega_1^2=\beta^2 d\varphi^2$. All $\theta_i$
range from 0 to $\pi$, while $\varphi$ goes from 0 to $2\pi$. The
constants $C$ and $R_0$ can be determined as

\be C^2 = {8\over 3(n+2)}\Lambda - {4(n-1)\over 3(n+2)}E^2
\;\;\;,\;\;\; R_0^{-2} = {2\over (n-1)(n+2)}\left(\Lambda +
{3\over 2}E^2 \right).\label{ctes-n-dim}\ee

\noindent If we want flat 4D space we have to tune $E = \sqrt{{2
\Lambda \over n-1}}$. It is interesting to note that we can tune
the effective cosmological constant to zero using the four form
v.e.v., that is not a parameter in the Lagrangian, but a dynamical
variable \cite{Turok:1998he} (see also \cite{Bousso:2000xa}).

 Notice that in the definition of $d\Omega_1^2$ we
allowed for an arbitrary parameter $\beta$ (a deficit angle in
$\varphi$). Values of this parameter different from one will be
associated with the presence of a $(n+1)$-brane at the points with
$sin^2\theta_1=0$ with nonzero tension, as we show in the next
subsection.

\subsection{Introducing ($n$+1)-branes.}

To show that the only effect of a ($n$+1)-brane with nonzero
tension in this background will be to induce a conical singularity
in the two dimensional transverse space, without affecting the 4D
Hubble expansion, we will follow \cite{Navarro:2003vw}, and we
will consider a regularization of a $\delta$-like brane placed at
the points where $sin(\theta_1)=0$\footnote{Alternatively we could
compute the the Einstein tensor for the metric (\ref{sol-n-dim}),
along the lines of \cite{Carroll:2003db,Kogan:2001yr}, and show
that it has a term proportional to $\delta(sin(\theta_1))$, that
would be cancelled if we consider the contribution to the
energy-momentum tensor of a $(n+1)$-brane located at
$\theta_1=0,\pi$ with tension proportional to $(1-\beta)$.}. We
will solve Einstein equations in all the spacetime (that now will
not have any singularity) and we will take the infinitely thin
limit to obtain the relation between the deficit angle and the
brane tension. Notice that for $n>2$, the ($d-2$)-dimensional
submanifold defined as the points in which $sin(\theta_1)=0$ is
connected, so we have to introduce a single brane in this
background, while for $n=2$ we had to introduce two 3-branes (at
the northern and southern poles of the sphere)
\cite{Carroll:2003db,Navarro:2003vw}. To be able to introduce a
single brane when $n=2$ it is necessary to consider
compactifications in a disk \cite{Navarro:2003vw}. The reason for
this is that the points with $sin(\theta_1)=0$ form the manifold
${\cal M}_4\times S^{n-2}$, with ${\cal M}_4$ being 4D Minkowski
or $dS_4$, and all $S^n$ are connected except for $S^0$, that
consists of just two disconnected points (if we define $S^n$ as
the set of points in $R^{n+1}$ at unit distance from the origin).

So we add a term to the energy momentum tensor, $\Delta
T_{MN}(\theta_1,\epsilon)$, such that it is finite everywhere
(when $\epsilon>0$), zero for $sin(\theta_1)>\epsilon$ and

\be lim_{\epsilon \rightarrow 0} \Delta T_{M N}(\theta_1,\epsilon)
=
-\left(\begin{array}{cc}g_{a b}{T_0\over 2\pi \sqrt{\kappa}}\delta(sin(\theta_1)) & \\
& 0 \end{array}\right) \label{limit},\ee

\noindent where the indices $a,b$ run over all the dimensions
except $\theta_1$ and $\varphi$, and $\kappa$ is the determinant
of the $2\times 2$ metric in the two dimensional transverse space
spanned by $(\theta_1,\varphi)$. In this way the energy momentum
tensor in the right hand side of eq.(\ref{limit}) would be derived
from a term in the action given by

\be \Delta S = \int d^dx\sqrt{|g|}{T_0\over 2\pi
\sqrt{\kappa}}\;\delta \left(sin(\theta_1)\right),
\label{brane}\ee

\noindent that is, a $\delta$-like ($n$+1)-brane source (see the
appendix of \cite{Leblond:2001xr} for a discussion of delta
functions in curved manifolds). When $sin(\theta_1)>\epsilon$ the
solution is given by eqs.(\ref{sol-n-dim},\ref{ctes-n-dim}), while
for $sin(\theta_1)<\epsilon$ we will consider solutions of the
form

\be ds^2 = dt^2 - e^{Ct}d\vec{{\bf x}}^2 - R^2_0
d\tilde{\Omega}_n^2 \label{sol-n-dim-rho} \ee

\noindent where $d\tilde{\Omega}_n^2$ is defined again as

\be d\tilde{\Omega}_n^2 = d\theta^2_{n-1} + sin^2 \theta_{n-1}
d\tilde{\Omega}_{n-1}^2 \ee

\noindent but now $d\tilde{\Omega}_2^2=d\theta_1^2 +
\rho(\theta_1)^2d\varphi^2$. The constants $C$ and $R_0$ are given
again by eq.(\ref{ctes-n-dim}). It is always possible to consider
regularizations of the brane such that one finds solutions of the
form (\ref{sol-n-dim-rho}) for some $\rho(\theta_1)$. Einstein
equations for this ansatz can be written as

\be - \left(\begin{array}{cc}g_{a b}{1\over \sqrt{\kappa}}\left(\rho(\theta_1)+\rho''(\theta_1)\right) & \\
& 0 \end{array}\right)= \Delta T_{MN}(\theta_1,\epsilon).\ee

\noindent It is clear that when $\Delta T_{MN}=0$ the metric in
eq.(\ref{sol-n-dim}) ($i.e.$ $\rho(\theta_1)=\beta sin(\theta_1)$)
is a solution. To obtain the matching condition that relates the
brane tension with the deficit angle we integrate the previous
equation in $\theta_1$ and take the limit $\epsilon \rightarrow
0$. For doing this we do not need to assume any particular
regularization of the brane, since in the thin limit the result
will be independent of the regularization used. We only have to
take into account that $\rho(0)=0$, $\rho'(0)=1$ (any other value
would produce a conical singularity and we are assuming that the
regularized solution is regular everywhere) and when
$sin(\theta_1)>\epsilon$ the solution is given by
eq.(\ref{sol-n-dim}). Doing this and taking the thin limit we get

\be 1-\beta={T_0\over 2\pi}.\label{defang} \ee

The brane will have four noncompact dimensions and $n-2$
dimensions compactified in $S^{n-2}$ with a radius given by
eq.(\ref{ctes-n-dim}). This implies that when $n>2$, since
Standard Model fields should be able to propagate in all the
dimensions of the brane, $R_0$ has to be small, of the order of
$TeV^{-1}$, to avoid light KK states of observable particles. A
solution to the hierarchy problem making use of large extra
dimensions like in \cite{Arkani-Hamed:1998rs} is only possible in
this scenario when $n$=2.

\section{Conclusions}

In this paper we continued the exploration of the scenarios
proposed in \cite{Carroll:2003db,Navarro:2003vw}. We have shown
that if a monopole like configuration of a $U(1)$ gauge field is
used to compactify the two extra dimensions in a sphere like in
\cite{Carroll:2003db,Randjbar-Daemi:1982hi}, the fine tuning
required in order to obtain a small 4D vacuum energy involves
brane parameters, spoiling the nice features of the model. If a
four form field strength is used instead, the required tuning is
completely independent of the brane tension, and so is the
effective 4D cosmological constant. It would be very interesting
to embed this scenario in a supersymmetric theory, since the
required tuning could be the consequence of supersymmetry, like in
\cite{Salam:1984cj,Aghababaie:2002be}\footnote{As we have
commented previously, once we introduce a brane with nonzero
tension in this model, solutions with flat 4D space are no longer
preserved, for reasons similar to those explained in section 2,
since a monopole configuration of a $U(1)$ gauge field is used to
obtain the spontaneous compactification in $S^2$.}.

We also generalized the solution to an arbitrary number of
dimensions, keeping four noncompact dimensions and compacifying
the rest in $S^{n}$. We have seen that it is also possible to
introduce a $(n+1)$-brane with arbitrary tension in these
backgrounds, and the effective 4D cosmological constant is not
affected by this tension.

In this letter we have not addressed the important issue of the
stability of the solutions. A complete analysis of the spectrum of
fluctuations is beyond the scope of the present work, but we will
make some remarks about this subject. It is generally nontrivial
to check whether a spontaneously compacified solution is stable or
not in general relativity
\cite{Cline:2003ak,Carroll:2001ih,Gunther:2003zn,DeWolfe:2001nz,Bousso:2002fi}.
Recently an analysis of the fluctuation spectrum of the model
presented here for $n=2$ was carried out in \cite{Cline:2003ak},
where if was concluded that the model is stable. However, it has
been shown  in \cite{Bousso:2002fi} that this kind of solutions
can be unstable for certain values of the gauge flux and certain
dimensionalities. In any case, we regard these constructions as
intermediate steps towards finding a more complete theory that
addresses the problems left here. For instance, as we have said
before, it would be desirable to embed these solutions in
supergravity, since this is the most promising way to cancel the
bulk contributions to the effective 4D cosmological constant and
this could change the stability properties of the solutions (one
could argue that for the better). Also, a new possible source of
instability when $n>2$ corresponds to deformations that shrink the
size of the extra dimensions of the brane to zero (this is always
possible since we know from homotopy theory that $\pi_{n-2}(S^n)$
is trivial). In any particular model it would be necessary to
check that the modes associated with these deformations have a
positive mass squared.

\section*{Acknowledgments} We thank Sacha Davidson and Jose Santiago for conversations.

\section*{Note Added}

After this work was finished ref.\cite{new} appeared, where the
introduction of branes in the SUSY model of
\cite{Salam:1984cj,Aghababaie:2002be} was considered. In this
reference it was also noticed that the introduction of branes
affects the solution and generates an effective 4D cosmological
constant proportional to the deficit angle in the simplest
realization of the model due to the quantization condition of the
gauge field (section 4.2). However, we have shown here that this
problem is absent if we consider a four form field strength
instead of the monopole to spontaneously compactify the extra
dimensions in a sphere.



\begin{thebibliography}{99}



\bibitem{Arkani-Hamed:1998rs}
N.~Arkani-Hamed, S.~Dimopoulos and G.~R.~Dvali,
Phys.\ Lett.\ B {\bf 429}, 263 (1998) [arXiv:hep-ph/9803315].
\bibitem{Randall:1999ee}
L.~Randall and R.~Sundrum,
Phys.\ Rev.\ Lett.\  {\bf 83} (1999) 3370 [arXiv:hep-ph/9905221].
\bibitem{Arkani-Hamed:2000eg}
N.~Arkani-Hamed, S.~Dimopoulos, N.~Kaloper and R.~Sundrum,
Phys.\ Lett.\ B {\bf 480} (2000) 193 [arXiv:hep-th/0001197].
\bibitem{Kachru:2000hf}
S.~Kachru, M.~B.~Schulz and E.~Silverstein,
Phys.\ Rev.\ D {\bf 62} (2000) 045021 [arXiv:hep-th/0001206].
\bibitem{Csaki:2000wz}
C.~Csaki, J.~Erlich, C.~Grojean and T.~J.~Hollowood,
Nucl.\ Phys.\ B {\bf 584} (2000) 359 [arXiv:hep-th/0004133].
\bibitem{Forste:2000ft}
S.~Forste, Z.~Lalak, S.~Lavignac and H.~P.~Nilles,
JHEP {\bf 0009} (2000) 034 [arXiv:hep-th/0006139].
\bibitem{Carroll:2003db}
S.~M.~Carroll and M.~M.~Guica,
arXiv:hep-th/0302067.
\bibitem{Navarro:2003vw}
I.~Navarro,
arXiv:hep-th/0302129.
\bibitem{Chen:2000at}
J.~W.~Chen, M.~A.~Luty and E.~Ponton,
JHEP {\bf 0009} (2000) 012 [arXiv:hep-th/0003067].
\bibitem{Cline:2003ak}
J.~M.~Cline, J.~Descheneau, M.~Giovannini and J.~Vinet,
arXiv:hep-th/0304147.
\bibitem{Randjbar-Daemi:1982hi}
S.~Randjbar-Daemi, A.~Salam and J.~Strathdee,
Nucl.\ Phys.\ B {\bf 214} (1983) 491.
\bibitem{Freund:1980xh}
P.~G.~Freund and M.~A.~Rubin,
Phys.\ Lett.\ B {\bf 97} (1980) 233.
\bibitem{Salam:1984cj}
A.~Salam and E.~Sezgin,
Phys.\ Lett.\ B {\bf 147} (1984) 47.
\bibitem{Aghababaie:2002be}
Y.~Aghababaie, C.~P.~Burgess, S.~L.~Parameswaran and F.~Quevedo,
arXiv:hep-th/0212091.
\bibitem{Turok:1998he}
N.~Turok and S.~W.~Hawking,
Phys.\ Lett.\ B {\bf 432} (1998) 271 [arXiv:hep-th/9803156].
\bibitem{Bousso:2000xa}
R.~Bousso and J.~Polchinski,
JHEP {\bf 0006} (2000) 006 [arXiv:hep-th/0004134].
\bibitem{Kogan:2001yr}
I.~I.~Kogan, S.~Mouslopoulos, A.~Papazoglou and G.~G.~Ross,
Phys.\ Rev.\ D {\bf 64} (2001) 124014 [arXiv:hep-th/0107086].
\bibitem{Leblond:2001xr}
F.~Leblond, R.~C.~Myers and D.~J.~Winters,
JHEP {\bf 0107} (2001) 031 [arXiv:hep-th/0106140].
\bibitem{Carroll:2001ih}
S.~M.~Carroll, J.~Geddes, M.~B.~Hoffman and R.~M.~Wald,
Phys.\ Rev.\ D {\bf 66} (2002) 024036 [arXiv:hep-th/0110149].
\bibitem{Gunther:2003zn}
U.~Gunther, P.~Moniz and A.~Zhuk,
arXiv:hep-th/0303023.
\bibitem{DeWolfe:2001nz}
O.~DeWolfe, D.~Z.~Freedman, S.~S.~Gubser, G.~T.~Horowitz and
I.~Mitra,
Phys.\ Rev.\ D {\bf 65} (2002) 064033 [arXiv:hep-th/0105047].
\bibitem{Bousso:2002fi}
R.~Bousso, O.~DeWolfe and R.~C.~Myers,
arXiv:hep-th/0205080.
\bibitem{new}
Y.~Aghababaie, C.~P.~Burgess, S.~L.~Parameswaran and F.~Quevedo,
arXiv:hep-th/0304256.







\end{thebibliography}
\end{document}